\documentclass[onecollarge,natbib]{svjour2}
\bibpunct{[}{]}{;}{n}{}{,} % to get "[numbered]" references from natbib
\smartqed  % flush right qed marks, e.g. at end of proof
\usepackage{graphicx,color}
\usepackage[normalem]{ulem}

\newcommand{\cd}{\makebox[0.08cm]{$\cdot$}}
\journalname{Few-Body Systems (EFB22)}
\begin{document}

\title{Transition electromagnetic  form factor in the Minkowski space  Bethe-Salpeter approach%\thanks{Grants or other notes
}


\author{J. Carbonell         \and
        V.A. Karmanov %etc.
}


\institute{J. Carbonell \at
              Institut de Physique Nucl\'eaire,
Universit\'e Paris-Sud, IN2P3-CNRS, 91406 Orsay Cedex, France \\
%              Tel.: +123-45-678910\\
%              Fax: +123-45-678910\\
              \email{carbonell@inpo.in2p3.fr}           %  \\
           \and
           V.A. Karmanov \at
            Lebedev Physical Institute, Leninsky Prospekt 53,
119991 Moscow, Russia\\
\email{karmanov@sci.lebedev.ru}
}

\date{Received: date / Accepted: date}

\maketitle

\begin{abstract}
Using the solutions of the Bethe-Salpeter equation in Minkowski space for  bound and scattering states found in previous works, we
calculate the transition electromagnetic form factor describing the electro-disintegration of a bound system. 
\end{abstract}

\section{Introduction}
\label{intro}
During the last few years there has been a renaissance of the  Bethe-Salpeter (BS) approach \cite{bs} in its original Minkowski space formulation.
It is caused by the progress in finding new methods which allow to overcome the difficulties resulting 
from the singularities of the BS equation and to obtain solution in the Minkowski space both on- and off-mass shell. 
The on-shell solution, found  few decades ago \cite{tjon,Schwartz_Morris_Haymaker},  allows to compute the nucleon-nucleon (NN) scattering phase shifts and  parametrize the NN-interaction. 
The off-shell solution for the bound \cite{bsCK} and scattering states \cite{KC_LCM_Cracovie_2012,kc_FB20,CK_Baldin_2012,CK_PLB_2013} is much more recent
and allows to properly calculate the electromagnetic  elastic  \cite{ckm_epja,ck-trento}
and  transition form factors describing the electro-disintegration of  a bound system (e.g. the deuteron). 
Note that the Euclidean  solution cannot be used for that purpose, since the Wick rotation can be done in the BS equation itself, 
but not in the integral expression giving the form factors \cite{ck-trento}.

The full BS program has been achieved for a  NN system interacting via separable kernel \cite{burov} 
since in this case, the singular integrals -- both in the equation and in the form factors -- are performed analytically.
However it is not yet fully realized  for a field-theoretically inspired description. 
For one-boson exchange kernel the analytical treatment of singularities is also possible due to the fact that the bound state BS solution is found in the form of the Nakanishi integral representation. In this case one can also calculate the integrals analytically and express the form factor \cite{ckm_epja} through the non-singular Nakanishi weight function \cite{nak63}. 
For the scattering states the development of this method (based on the Nakanishi representation  \cite{nak63}) is in progress \cite{fsv-2012} but  did not yet enter in the stage of numerical calculations. 
On the other hand, we have recently developed a different approach \cite{KC_LCM_Cracovie_2012,kc_FB20,CK_Baldin_2012,CK_PLB_2013}
based on the direct treatment of the singularities and which  allows to find both the bound and scattering state solutions. 

The direct treatment of singularities is required also when calculating the transition form factors. 
In the present work we briefly explain our approach to this problem and present the first numerical results. 
In absence, for the time being, of  scattering state solutions in the Nakanishi form, our method is the only one that allows to  calculate this quantity. 
We consider in this contribution only the case of spinless particles. The non-zero spin case complicates the algebra but do not change the singularities of the propagators.

%%%%%%%%%%%%%%%%%%%%
\section{Calculating form factor}
\label{sec:1}

\vspace{-0.5cm}
\begin{figure}[h!]
\begin{center}
\includegraphics[width=0.3\textwidth]{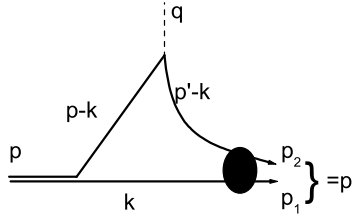}
\caption{Feynman diagram for the EM transition form factor.}\label{triangle}
\end{center}
\end{figure}

\vspace{-0.5cm}
Let us start with the expression for the transition current, represented graphically  in figure \ref{triangle}. 
\begin{equation}\label{ffin0}
\tilde{J}_{\mu}=i\int \frac{d^4k}{(2\pi)^4}\,
\frac{(p_{\mu}+p'_{\mu}-2k_{\mu})\Gamma_i \left(\frac{1}{2}p -k,p\right)\Gamma_f
\left(\frac{1}{2}p'-k,p'\right)} {(k^2-m^2+i\epsilon)[(p-k)^2-m^2+i\epsilon]
[(p'-k)^2-m^2+i\epsilon]}.
\end{equation}
The plane wave contribution will  not be considered in this short presentation.

The quantities  $\Gamma_i$ and $\Gamma_f$ are respectively  the initial (bound state) and  final (scattering) vertex functions.
 $\Gamma_f$   is the half-off-shell S-wave scattering BS amplitude found in \cite{{kc_FB20}}. 
Like any scattering amplitude,  $\Gamma_f$ depends on the initial and final four-momenta satisfying the conservation law. 
Therefore, the expression (\ref{ffin0}) depends on the four-momenta $p,p_1,p_2$ (excluding $q$ from the conservation law) or alternativley on 
$p,p'=p_1+p_2$ and $p_s=(p_1-p_2)/2$.
For a given partial wave, say S-wave, we average it over the angles of the vector $\vec{p}_s$ in the c.m. frame of  the final state where $\vec{p'}=0$. 
After that the four-vector $p_s$ does not enter in the decomposition of $\tilde{J}_{\mu}$, 
though the result depends on the module
 $|\vec{p}_s|$ which we will still denote as $p_s$. It determines the final state mass $M_f=2\sqrt{m^2+p_s^2}$.
Therefore we have in our disposal the four-vectors  $p,p'$ only.

Notice also that in the approximation given by fig. \ref{triangle}, restricted to the one-body electromagnetic current for the interaction of photons with constituents, the current
$\tilde{J}_{\mu}$ is not conserved, that is  $(p'-p)_{\mu}\tilde{J}_{\mu} \neq 0$.
It can be decomposed as:
\begin{equation}\label{ffa}
\tilde{J}_{\mu}=(p_{\mu}+p'_{\mu})F(Q^2)+ (p'_{\mu}-p_{\mu})F'(Q^2)
\end{equation}
In this situation, one can chose one of the two following strategies:
({\it i})~find both $F$ and $F'$ and  use the current (\ref{ffa}), in spite of its non-conservation, to calculate the electro-disintegration cross section;
({\it ii})~start with (\ref{ffa}) and  construct the conserved current
\begin{equation}\label{ffb}
J_{\mu}=\tilde{J}_{\mu}-\frac{q_{\mu}}{q^2}(q\cd\tilde{J})=\left[(p_{\mu}+p'_{\mu})+ \frac{({M_f}^2-M_i^2)}{Q^2}(p'_{\mu}-p_{\mu})\right]F(Q^2)
\end{equation}
which satisfies to $q\cd J=0$ ($q=p'-p$, $q^2=(p'-p)^2$, $Q^2=-q^2$). Due to the constraint $q\cd J=0$, the current $J_{\mu}$ is determined only by the form factor $F(Q^2)$, 
and this is the quantity that will be calculated below. The form factor $F'(Q^2)$,  if at all necessary, can be obtained analogously.

It is convenient to carry out the calculations in the system of reference where $p'_0=p_0$ (i.e. $q_0=0$) and $\vec{p}$ and $\vec{p'}$ are collinear (i.e., they are either parallel or anti-parallel to each other, depending on the $Q^2$ value). 
In the elastic case, this system coincides with the Breit frame $\vec{p}+\vec{p'}=0$, where $|\vec{p}|=|\vec{p'}|$ and  $p'_0=p_0$. 
In the inelastic case, since one still has  $p'_0=p_0$,  the three-momenta are different $|\vec{p}|\neq|\vec{p'}|$.  
This frame can be obtained by performing a Lorentz transform from the Breit frame.

Using this reference frame,  the form factor $F(Q^2)$ can be computed by taking the zero component of the current:
$$
J_{0}=\tilde{J}_{0}=2p_{0}F(Q^2)
$$
That is:
\begin{equation}\label{ffin1}
F(Q^2)=\frac{i}{p_0}\int \frac{d^4k}{(2\pi)^4}\,
\frac{(p_0-k_0)\Gamma_i \left(\frac{1}{2}p -k,p\right)\Gamma_f
\left(\frac{1}{2}p'-k,p'\right)} {(k_0^2-E_{\vec{k}}^2+i\epsilon)[(p_0-k_0)^2-E_{\vec{p}-\vec{k}}^2+i\epsilon]
[(p'_0-k_0)^2-E_{\vec{p'}-\vec{k}}^2+i\epsilon]},
\end{equation}
where  $E_{\vec{k}}=\sqrt{m^2+\vec{k}^2}$  is the on-shell energy  (and similarly for other energies).

This 4D integral  is in practice a 3D one, since the azimuthal integration gives simply a factor $2\pi$.
To calculate it numerically, we treat the singularities in a similar way to what was done for solving the scattering states BS  equation \cite{kc_FB20}. 
 Namely, we represent each of three propagators as the sum  of the principal value (PV) integrals and the $\delta$-function contributions,
that is, symbolically:
\begin{equation}\label{PVdelta}
F(Q^2)=\int (PV+i\pi\delta)(PV+i\pi\delta)(PV+i\pi\delta)\ldots
\end{equation} 

The non-zero contributions result from the terms containing two, one and zero $\delta$-functions.
Performing explicit integrations of the $\delta$-functions, we get, correspondingly, the sum of 
1D, 2D and 3D PV integrals. To treat the pole singularities in the $k_0$ variable in the PV integrals, we introduce a function $g(k_0)$ which has the same pole singularities -- and  only them -- and the same residues as the  integrand of (\ref{ffin1}),  once  (\ref{PVdelta}) being  inserted. 
We perform  usual subtraction technique in the integrand  with $g(k_0)$  in such a way that the difference is smooth function and can be safely integrated numerically.  
Since the  function $g(k_0)$  is the sum of a few pole terms only, its integral  is  calculated analytically. 
All other singularities (the vertex functions $\Gamma$, or in $k$ variable, after integration over $k_0$) are logarithmic  or weaker and therefore can be also integrated numerically, though with some care. 
In this way we can properly treat the singularities in the integral (\ref{ffin1}) and calculate the transition form factor. The details of this calculation will be given in
a subsequent  publication \cite{KC_2014}.

As a test of our calculation we have considered the case $\Gamma_i = \Gamma_f =1$.
We used the  two-parameters  Feynman parametrization to calculate analytically the 4D integral  (\ref{ffin0}). 
The resulting integral is calculated analytically over one Feynman parameter  and numerically over the second one. We took small value of $\epsilon$ in the denominator of (\ref{ffin0})  and obtained a stable result relative to variation of $\epsilon$.
In this way, we reproduced both real and imaginary parts of the transition form factor vs. $Q^2$ found by the method described above.
Taking $M_f=M_i$, we reproduced the (real) elastic form factor. 

\bigskip

\begin{figure}[h!]
\begin{center}
\begin{minipage}[t]{75mm}
\begin{center}
\includegraphics[width=0.9\textwidth]{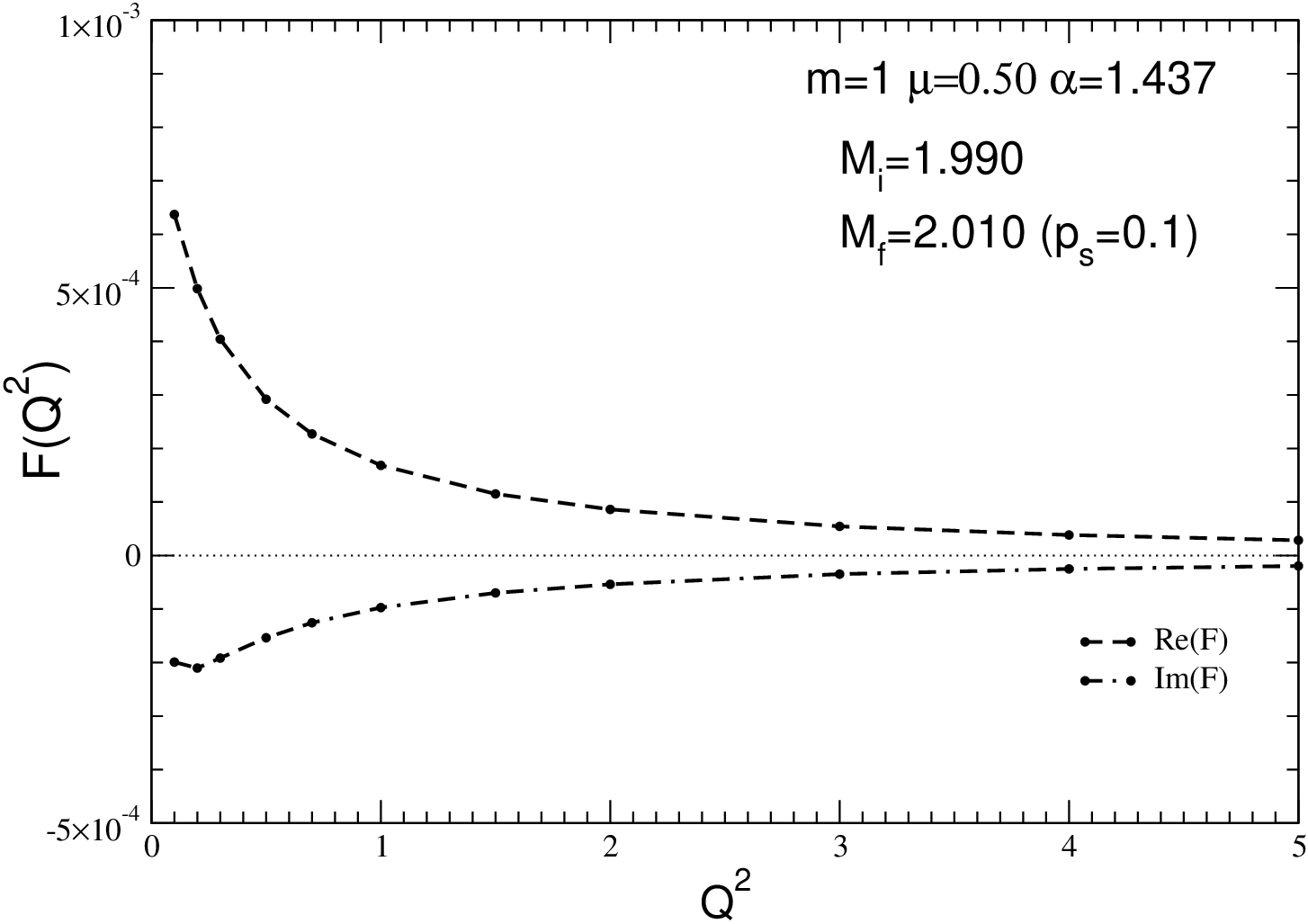}
\caption{Real  (dashed) and imaginary (dot-dashed) parts  of the transition form factor $F(Q^2)$   for $p_s=0.1$.} \label{fig2}
\end{center}
\end{minipage}
\hspace{0.3cm}
\begin{minipage}[t]{75mm}
\begin{center}
\includegraphics[width=0.9\textwidth]{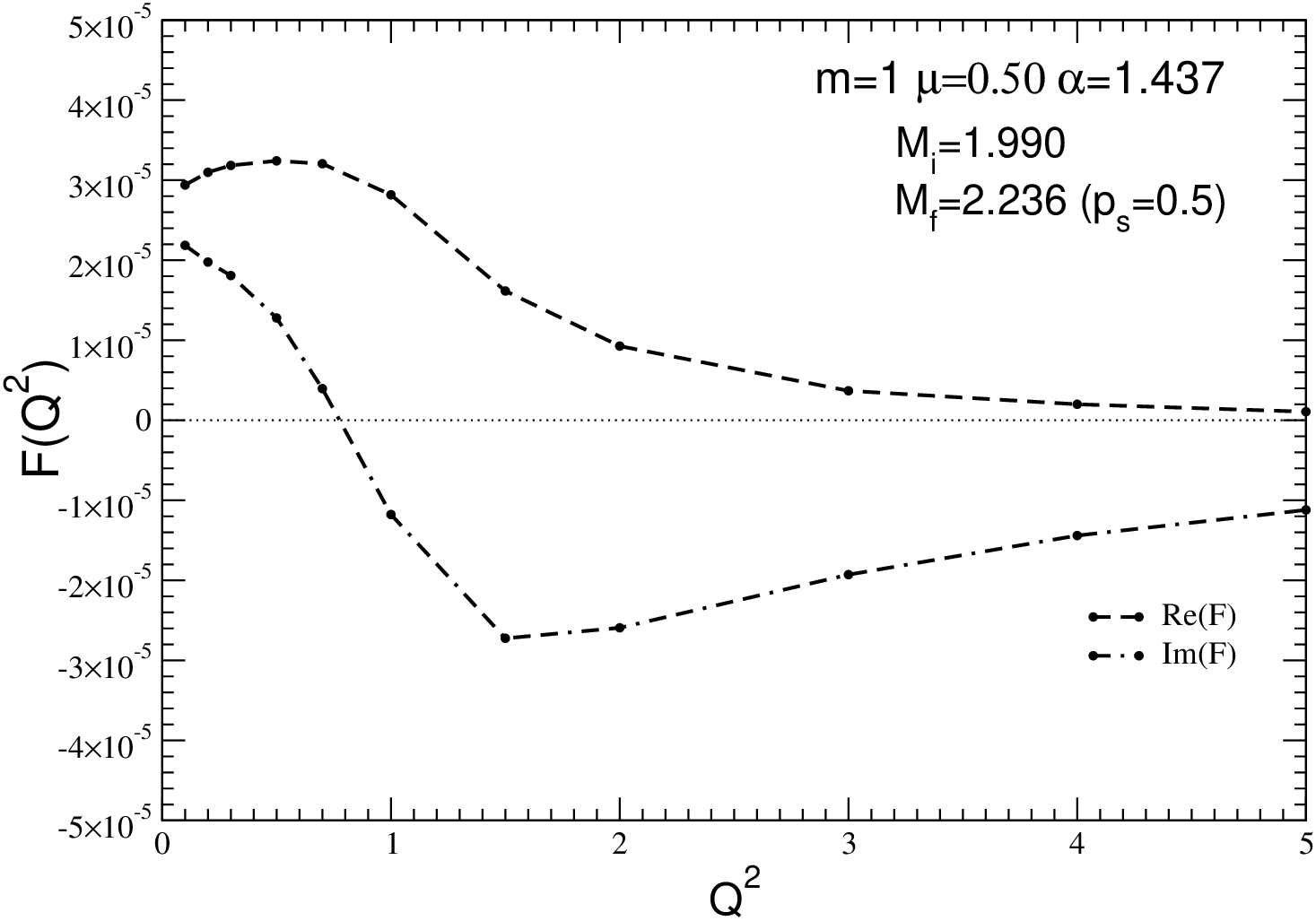}
\caption{The same as in fig. \protect{\ref{fig2}}  for $p_s=0.5$.}\label{fig3}
\end{center}
\end{minipage}
\end{center}
\end{figure}

\section{Numerical results}

We present the results corresponding to the one-boson exchange model with the constituent mass $m=1$, exchange boson  mass $\mu=0.5$ and the coupling constant $\alpha=1.437$, providing a bound state with the mass  $M_i=1.99$.
The initial bound state BS amplitude $\Gamma_i$ and the final scattering state amplitudes $\Gamma_f$ have been obtained by the method
described in \cite{kc_FB20}.
Using these solutions, the transition form factor $F(Q^2)$
was calculated by the methods presented above. Its real and imaginary parts  vs. $Q^2$  for the parameters
 $M_i=1.99$ and $M_f=2.01$  ($p_s=0.1$) are shown in fig. \ref{fig2}. 
For these kinematical parameters the form factor  is almost real (like the elastic form factor) since the final mass is very close to the initial one.

For $p_s=0.5$ ($M_f=2\sqrt{p_s^2+m^2}=2.236$) and for the same values of other parameters the transition form factor is shown in fig.  \ref{fig3}.
Now, for considerably larger inelasticity (i.e., for larger effective final state mass), the imaginary part of form factor is comparable with its real part. 
In our calculations, we have no restrictions for the values of momentum transfer $Q^2$ and the final state mass $M_f$. Several aspects remain to be clarified, like e.g.  the relative importance of the plane wave graph and
the dependence on the binding energy of the initial state.

%%%%%%%%%%%%%%%%%%%%%%%%%%%%%%%%%%%%%%%%%%%%
\section{Conclusion}

The solutions of Bethe-Salpeter equation in the natural Minkowski metric
are difficult by the singularities presented in the integral equation and in the solution itself.
The euclidean solutions are much more easy to get but they are
not suitable for obtaining the scattering amplitudes as well as the electromagnetic and/or transition form factors
due to the impossibility to perform the Wick rotation in the corresponding integral expressions.

A series of works have been undertaken in the last years to overcome this difficulty.
They had the aim of safely obtaining  the Minkowski solutions for bound and scattering states 
\cite{KC_LCM_Cracovie_2012,kc_FB20,CK_Baldin_2012,CK_PLB_2013,fsv-2012}
and compute the elastic electromagnetic form factor \cite{ckm_epja,ck-trento}.
This contribution is inserted in this effort and presents the  first results on the transition electromagnetic form factor.

\end{document}